\documentstyle[psfig,emulateapj]{article}

\def\arcsec{\hbox{$^{\prime\prime}$}}

\def\farcs{\hbox{$.\!\!^{\prime\prime}$}}         
\def\gsim{\mathrel{\hbox{\rlap{\lower.55ex \hbox {$\sim$}}
                   \kern-.3em \raise.4ex \hbox{$>$}}}}

\begin{document}

\title{The discovery of polarization
       in the afterglow of GRB\,990510 with the ESO Very Large Telescope}

\author{
R.A.M.J. Wijers\altaffilmark{1}, 
P.M. Vreeswijk\altaffilmark{2}, 
T.J. Galama\altaffilmark{2},
E. Rol\altaffilmark{2},
J. van Paradijs\altaffilmark{2,3}, 
C. Kouveliotou\altaffilmark{4,5}, 
T. Giblin\altaffilmark{3},
N. Masetti\altaffilmark{6}, 
E. Palazzi\altaffilmark{6},
E. Pian\altaffilmark{6}, 
F. Frontera\altaffilmark{6,7}, 
L. Nicastro\altaffilmark{8},
R. Falomo\altaffilmark{9},
P. Soffitta\altaffilmark{10},
L. Piro\altaffilmark{10}
}

\altaffiltext{1}{Dept.\ of Physics and Astronomy, State University of NY, Stony
Brook, NY 11794-3800, USA}

\altaffiltext{2}{Astronomical Institute `Anton Pannekoek', University
of Amsterdam, \& Center for High Energy Astrophysics, Kruislaan 403,
1098 SJ Amsterdam, The Netherlands} 

\altaffiltext{3}{Physics Department, University of Alabama in
Huntsville, Huntsville AL 35899, USA}

\altaffiltext{4}{Universities Space Research Association}

\altaffiltext{5}{NASA/MSFC, Code ES-84, Huntsville AL 35812, USA}

\altaffiltext{6}{Istituto Tecnologie e Studio Radiazioni
Extraterrestri (TESRE), CNR, Via P. Gobetti 101, 40 129 Bologna,
Italy}

\altaffiltext{7}{Physics Department, University of Ferrara, Via Paradiso, 
     12, 44100 Ferrara, Italy}

\altaffiltext{8}{IFCAI, CNR, Via Ugo La Malfa 153, I-90146 Palermo, Italy}

\altaffiltext{9}{Osservatorio Astronomico di Padova, Vicolo dell'Osservatorio,
5, I-35122 Padua, Italy}

\altaffiltext{10}{Istituto Astrofisica Spaziale, CNR, 
    Via Fosso del Cavaliere, 00133 Roma, Italy}

\begin{abstract}
Following a BeppoSAX alert (Piro 1999a) and the discovery of the
OT at SAAO (Vreeswijk et~al.\ 1999a), we observed GRB\,990510 with the
FORS instrument on ESO's VLT Unit 1 (`Antu').
The burst is unremarkable in gamma
rays, but in optical is the first one to show good evidence for jet-like
outflow (Stanek et al.\ 1999, Harrison et~al.\ 1999).  We report the
detection of significant linear polarization in the afterglow:
it is $1.6\pm0.2$\% 0.86 days after trigger, and after 1.81 days is
consistent with that same value, but much more uncertain.
The polarization angle is constant on a time scale of hours, and may be
constant over one day.
We conclude that the polarization is intrinsic to the source and due
to the synchrotron nature of the emission, and discuss the random and
ordered field geometries that may be responsible for it.
\end{abstract}

\section{Introduction
         \label{intro}
	 }

It is now well established that gamma-ray burst afterglows are the result
of relativistic blast waves (Rees and M\'esz\'aros 1992, M\'esz\'aros
and Rees 1997, Wijers, Rees, and M\'esz\'aros 1997, Sari, Piran and
Narayan 1998; see Piran 1999 for a review)
emitting primarily synchrotron radiation (Galama et~al.\
1998a,b, Wijers and Galama 1999).  Synchrotron radiation is highly
polarized, with typical degrees of (linear) polarization for ordered
magnetic fields of $\sim$60\% (Hughes and Miller 1991) and one should
therefore not be surprised if GRB afterglows show a measurable amount
of polarization. If the shock takes place in a collimated outflow (jet)
one might expect, by analogy to what is observed for jets in AGNs,
degrees of linear polarization of $10-20$\% (Angel and Stockman 1980,
Muxlow and Garrington 1991). The strong intrinsic polarization of this
emission is lowered by averaging over the unresolved source (Gruzinov
and Waxman 1998, Gruzinov 1999, Medvedev and Loeb 1999, Loeb and Perna
1998), and thus far only an upper limit to afterglow polarization has
been set (Hjorth et~al.\ 1999).  Here we report the results of our
optical polarimetric observations of the afterglow of GRB~990510, one
and two days after trigger. We detect significant polarization on day
one, similar in magnitude and position angle to
the value obtained by Covino et~al.\ (1999) that same night.

\begin{figure}
\begin{minipage}{8.8cm}
  \psfig{figure=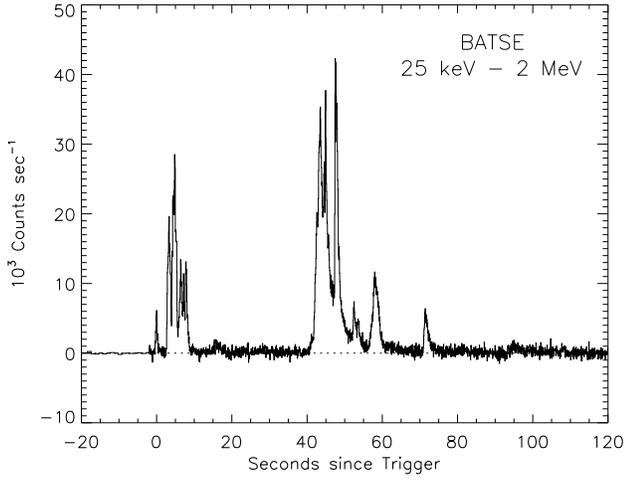,width=\textwidth}
  \caption[]{Time history of GRB\,990510 integrated
     over the four BATSE discriminator energy channels (25\,keV -- 2\,MeV) at
     64\,ms time resolution.
     \label{fig:batse}
}
\end{minipage}
\end{figure}

The prompt gamma-ray emission from GRB\,990510 was detected with BATSE
on Compton GRO on 1999 May 10.367 UT (Kippen et al. 1999), with Ulysses
(Hurley and Barthelmy 1999), and with the GRBM on BeppoSAX (Amati et~al.\
1999). The BATSE flux history (Fig.~\ref{fig:batse}) shows multiple peaks,
and a duration ($T_{90}$) of 68\,s.  The peak energy flux (25--2000 keV)
was  $(5.19 \pm 0.96) \times 10^{-6}$ erg cm$^{-2}$ s$^{-1}$, ranking
it in  the top $4\%$ among BATSE GRBs. The fluence is $(2.29 \pm 0.07)
\times 10^{-5}$\,erg\,cm$^{-2}$, placing it in the top $9\%$ of the BATSE
distribution. Assuming $H_0=70$ km s$^{-1}$ Mpc$^{-1}$, $\Omega_0=0.3$,
and $\Lambda=0$, we deduce a peak luminosity  $L_{\gamma} = 7.3 \times
10^{52}$ erg s$^{-1}$ and total  energy release $E_{\gamma} = 1.2 \times
10^{53}$\,erg (for $z=1.62$ and isotropic emission).  The time integrated
fit to the entire burst gives a peak energy (as defined in Band et~al.\
1993) $E_{p} = 147 \pm 4$\,keV, placing it in the center of the BATSE
$E_{p}$ distribution (Malozzi et al.\ 1995). The burst is therefore
unremarkable in gamma rays both w.r.t.\ the entire BATSE catalog and
w.r.t.\ other bursts with detected afterglows.

\begin{figure}
\begin{minipage}{8.8cm}
  \centerline{\psfig{figure=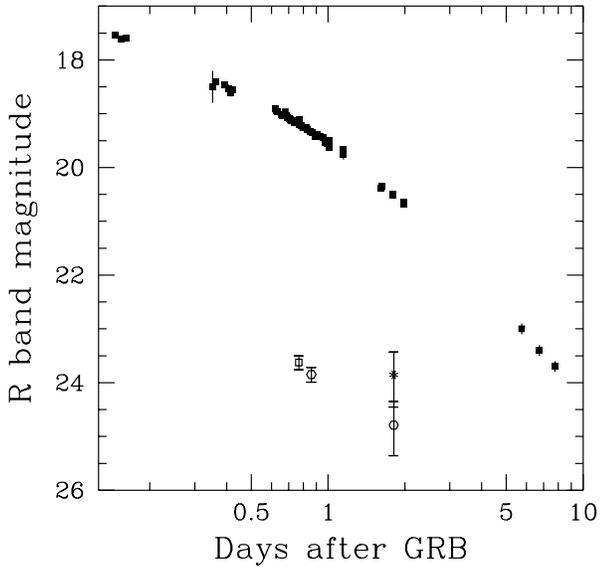,width=\textwidth}}
  \caption[]{The light curve of GRB\,990510 in $R$. Full symbols give the
           total flux (Axelrod et~al.\ 1999, Galama et~al.\ 1999,
           Stanek et~al.\ 1999, Vreeswijk et~al.\ 1999a, Covino et~al.\
           1999b,c, Bloom et~al.\ 1999, Lazzati et~al.\ 1999, Marconi et~al.\
           1999a,b) and show the break in
           the decay at about 1.5 days. Open symbols give the polarized flux
           in $R$, obtained by multiplying the total flux by the percentage
           polarization. square: polarization from
           Covino et~al.\ 1999; hexagons: our data,
           aperture photometry; star: our data, psf photometry (see text).
           \label{fig:rflux}
          }
\end{minipage}
\end{figure}

GRB\,990510 was located by the WFC on board BeppoSAX (Dadina et~al.  1999)
and its X-ray afterglow was detected by BeppoSAX as well (Kuulkers et~al.\
1999). The position of the WFC X-ray source is RA = 13h38m06s, DEC =
$-80^{\circ}$29$^{\prime}$.5 (equinox 2000.0), with an error radius of
3 arcminutes (Piro 1999a).  With the 1-m telescope at the South African
Astronomical Observatory (SAAO) we started imaging the error region at May
10.72, roughly 8.5 hours after the burst.  Comparison with the Digitized
Sky Survey revealed a previously unknown object at RA = 13h38m07.62s,
DEC = $-80^{\circ}29^{\prime}48.8^{\prime\prime}$ (Vreeswijk et
al. 1999a).  Following the identification we took low-resolution spectra
at the VLT of the optical transient (OT), setting a lower limit to the
redshift of $z=1.619\pm0.002$ through the identification of redshifted
absorption lines (Vreeswijk et al. 1999b).  Numerous photometric
observations show that the light curve is well--described by a power law
with a break occurring about 1.5 days after the burst (Stanek et al. 1999,
Harrison et~al.\ 1999; Fig.~\ref{fig:rflux}), which may be the result
of beaming (Rhoads 1999, M\'esz\'aros and Rees 1999, Sari et~al.\ 1999).

We report our polarimetric analysis and its results in Sect.~\ref{pola},
discuss possible interpretations in Sect.~\ref{orig}, and then 
summarize our findings.

\section{Polarimetric observations}
         \label{pola}

Optical polarization observations of GRB\,990510 were obtained with the
FOcal Reducer/low dispersion Spectrograph 1 (FORS1) on the European
Southern Observatory's (ESO) 8.2-m Antu telescope (VLT-UT1) on 1999
May 11.228 UT and May 12.17 UT. The polarization optics consist of a
phase retarder plate mosaic and a Wollaston prism.  A mask producing
20$^{\prime\prime}$ wide parallel strips was used to avoid overlap of
the ordinary and extraordinary components of incident light. The CCD
has 2k$\times$2k pixels of 0\farcs2 size.  Each observation consisted
of three Bessel R 10 minute exposures centered on the position of the
optical transient (Vreeswijk et al.  1999a). Each exposure was obtained
at a different phase retarder angle; we used a half wavelength plate
for the determination of the linear polarization. We also measured the
polarimetric standards BD--13$^{\circ}$5073 and BD--12$^{\circ}$5133
(Wagner and Szeifert 1999) on both nights.  For BD--13$^{\circ}$5073
we find $(P,\theta)=(4.90\pm0.08\%, 161.6\pm0.5^\circ)$, compared with
$(P,\theta)=(4.61\pm0.03\%,151.0\pm0.7^\circ)$ by Wagner and Szeifert, and
for BD--12$^{\circ}$5133 we find $(P,\theta)=(5.07\pm0.13\%,155.7\pm0.7)$,
compared with $(P,\theta)=(4.33\pm0.03\%,148.0\pm0.7^\circ)$.
Since the standard values were measured in $B$ band (Szeifert,
private communication), and $P$ is chromatic, we regard the agreement
as satisfactory. $\theta$ is hardly color-dependent, so the mean
offset between the standards and our measured values of $9.2\pm1.5$
degrees is real; all values quoted for the OT below are corrected
for this amount of instrumental polarization.
The typical seeing on May 11 and May 12 was 1\farcs0
and 2\farcs5, respectively. Details of the observations are given in
Table~\ref{tab:log}.

The CCD frames were bias subtracted and flat fielded with the NOAO IRAF
package in a standard way. The linear polarizations and the polarization
angles of the optical transient and 23 field stars were calculated
from each of the images using standard equations (Ramaprakash 1998). We
determined the Stokes parameters Q and U of the optical transient relative
to these field stars, which corrects for possible instrumental and
(local) interstellar polarization. No systematic variations of the
field star polarizations with position on the CCD or magnitude were
found. We therefore reference the OT Stokes parameters to the accurately
determined mean of the field stars, so the errors in the OT photometry
dominate the error in the polarization measurement.

On May 11 we measure $\bar{Q} = (1.05\pm0.04)$\%, $\bar{U} =
(-0.68\pm0.03)$\% for the weighted mean Stokes parameters of the field
stars (using aperture photometry). Correcting the measured $Q = (-0.28
\pm 0.18)$\% and $U = (-1.52\pm 0.26)$\% of the OT with these numbers
we find $Q_{\rm OT} = (-1.33 \pm 0.18)$\%, $U_{\rm OT} = (-0.84 \pm
0.26)$\%, corresponding to a linear polarization of (1.6 $\pm$ 0.2)\%
at a position angle $\theta = 98 \pm 5$ degrees.
Gaussian PSF-fitting photometry was also performed on the OT
and field stars using the DAOPHOT II package (Stetson 1987) and the {\sl
ALLSTAR\/} procedure in MIDAS. Combined with an alternative polarization
analysis method (di Serego Alighieri 1997), we find $P=(1.6\pm0.2)$\%
and $\theta=96\pm4$ degrees, in very good agreement with the result
obtained using aperture photometry.

Covino et al. (1999) found that on May 11 ($\sim$ 2 hours before
our observations) the afterglow of GRB~990510 showed linear
polarization at the level of (1.7 $\pm$ 0.2)\% with a position angle
$\theta = 101 \pm 3$ degrees, relative to the stars in the field
(values corrected from their preliminary report; 
Covino, private communication, and Covino et~al.\ 1999). 
So in two hours, the polarization shows no evidence of change.

On May 12 the optical transient was 1.2 magnitudes fainter ($R$=
20.65), and the observing conditions were much worse: overhead
cirrus and a seeing of 2\farcs5. We applied the same procedure to the
May 12 data, except that we used a smaller aperture and a fixed
position of the OT (from the previous data) to minimize the
contribution from a star 4\arcsec\ away.
We find for the comparison stars
$\bar{Q} = (1.06\pm0.04)$\% and $\bar{U} = (-0.38\pm0.04)$\%.
Correcting the measured $Q = (0.1 \pm 0.9)$\% and $U = (-2.0
\pm 1.2)$\% we find $Q_{\rm OT} = (-1.0 \pm 0.9)$\%, $U_{\rm OT} =
(-1.6 \pm 1.2)$\%. 
Since $U$ and $Q$ are determined from small differences in very high
signal-to-noise detections of the OT, their errors will be approximately
normally distributed. We therefore evaluated the mean value and 68\% confidence
interval for $P$ and $\theta$ by Monte Carlo drawing many
realizations of $U$ and $Q$, computing $P$ and $\theta$ for each, and 
inspecting the resulting distributions of $P$ and $\theta$. The
result is that $P=(2.2_{-0.9}^{+1.1})$\%
and $\theta =112_{-15}^{+17}$ degrees. This gives the 
impression that $P=0$ is fairly well excluded.
However, the simulations show that
if we had measured an unpolarized source with the same precision in $U$
and $Q$ we would have had an 11\% chance of measuring $P>2.2$\%, so
our detection is not very secure.
Using PSF photometry, we find for this data set $Q_{\rm OT} = (-2.5
\pm 2.6)$\% and $U_{\rm OT} =(-3.6\pm 2.6)$\%, consistent with the
aperture values, and resulting in $P=(5.2_{-2.2}^{+2.5})$\% and
$\theta =99_{-16}^{+19}$ degrees (and a 12\% chance probability).
The larger error is mostly due to the poorer
seeing in the presence of a nearby star. This leads to some problems
in the measurement that are not readily quantified as a random error,
so we consider our measurement on night 2 as tentative.
The polarized flux is plotted
in Fig.~\ref{fig:rflux}, along with the $R$ band light curve of the OT.

\section{Origin and implications of the polarization} 
         \label{orig}

Some (constant) polarization in the afterglow could be generated by
dust scattering by the host's interstellar medium.  For dust scattering
to polarize the light even by a few percent, at least that fraction
of the light must have been scattered.  This requires a path length
of many parsecs, which would cause a time delay of months between the
(scattered) polarized light and the direct light, and thus could not
cause polarization within a day of the GRB trigger.  Electron scattering
in the GRB itself could also lead to some polarization, as was seen,
e.g.\ in SN1998bw and attributed to asymmetries in the photosphere (Kay
et~al.\ 1998).  The degree of polarization could never be more than the
electron scattering optical depth, however, which is typically never
more than $10^{-6}$ after a day or so.

Intrinsic polarization is expected from any synchrotron source: $P_{\rm
max}\sim60$\% is normal from an emitting region with one direction of
the magnetic field. However, the net polarization from an unresolved
source will still be small if the direction of the polarization averages
out. There are two possible reasons why the polarization might average
out to nearly zero: highly tangled magnetic fields and very highly
symmetric field geometries. We now examine the consequences of both for
our measurements and their interpretation.

The magnetic field could be highly tangled, with only small-scale
structure, if it is generated by some form of turbulence. We can think
of this case as a source that consists of $N$ patches within which the
field has a single coherent direction, but no correlation between the
patches.  In that case, we expect a net polarization of order $P_{\rm
max}/\sqrt{N}$.  Gruzinov and Waxman (1998) considered a turbulently
generated magnetic field, which has such a small scale that it would
not likely leave a net polarization.  However, they suggested that the
coherence length of the field might grow, and the net polarization could
be a few to ten percent. Loeb and Perna (1998) suggested that microlensing
might amplify a few cells briefly, making the net polarization comparable
to the value for a single cell for a short time.

A more ordered field was discussed by Medvedev and Loeb (1999), who
consider the generation of a magnetic field parallel to the shock
front. Due to aberration, it would be parallel to the ring-like image
that the afterglow presents at late times (Panaitescu and M\'esz\'aros
1998, Sari 1998), causing a radial polarization. The mean polarization
of the image would still be zero for a spherical blast wave, due to
averaging over the unresolved ring.  Medvedev and Loeb (1999) suggest
that interstellar scintillation will cause polarization by selectively
magnifying part of the source; however this scintillation only occurs
at radio wavelengths, so it cannot explain the optical polarization.

If the symmetry of the emission itself is broken on the scale of
the source, significant net polarization will result (Gruzinov 1999).
There are good indications of asymmetry in GRB\,990510: the light curve in
optical steepens after about 1.5 days in a wavelength-independent manner
(Stanek et~al.\ 1999, Harrison et~al.\ 1999; Fig.~\ref{fig:rflux}).
Such a steepening would be well explained by beaming: when a burst is
caused by a jet-like outflow with opening angle $\theta$, the light
curve steepens around the time when the Lorentz factor of the jet
goes from $\Gamma>\theta^{-1}$ to $\Gamma<\theta^{-1}$ (Rhoads 1999,
M\'esz\'aros and Rees 1999).  Let our line of sight lie within the jet
cone, but away from the jet axis. As long as $\Gamma>>\theta^{-1}$ we see
the blast wave as if it were spherical, hence the polarization is zero.
As the jet slows down, we see the edge of the jet, causing asymmetry and
net polarization at about the same time as the light curve steepening
sets in. The direction of the polarization is constant, because it is
fixed by the geometry of the beam relative to our line of sight.  At late
times, the polarization will decrease again, because the emission opening
angle becomes much bigger than our offset from the center of the beam,
reducing the asymmetry.  So for polarization due to jets, we expect the
polarization be strongest near the time of the break in the light curve.
Note that because the effect of an ordered field could add up over the
source, as opposed to the random case, an ordered field need not be 
nearly as strong as the random field to dominate the net polarization.

The way to distinguish random- and ordered-field interpretations of
the polarization, then, is to look at the behavior of the polarization
angle. If it is constant, this argues in favor of an ordered field;
if it varies then the field is more likely to be random. Our data 
compared with the measurement of Covino et~al.\ (1999) shows a 
constant polarization for two hours.
Gruzinov and Waxman (1999) estimate that for the conditions
on night 1, i.e.\ a polarization of 1.6\% at 0.86 days since trigger,
the variation time scale of the polarization for a random field should
be about 0.25 days. This is sufficiently larger than the two-hour interval
in the data to make a random field consistent with the measured
constancy. For $N$ identical patches, each with an intrinsic
polarization of 60\%, we find that some 1100 patches will give an
expectation value for the net polarization equal to what we measure.
But the distribution of net polarizations is broad for any $N$, and
the 68\% likely range is $N=240-2600$. If the measurement on night 2 
is taken at face value, it means that the polarization angle is constant
over 1.0 days, rather longer than the predicted coherence time, and thus
that an ordered field is preferred over a turbulent one. Given the
problems with those data, we would rather consider this tentative
inference an illustration of what we can learn with present instrumentation,
under slightly more favorable conditions.

   \section{Conclusions
            \label{conc}
	   }

We have measured significant polarization in the afterglow of GRB\,990510,
which was an unremarkable burst in its gross gamma-ray and optical
properties, but notable for providing good evidence of beaming.
0.86 days after trigger, we find $(P=1.6\pm0.2)$\%, and a day later we
marginally detect polarization at a similar level.  We conclude that the
polarization is not due to interstellar or intra-source scattering and
attribute it to the synchrotron radiation from the blast wave itself.
The polarization is constant between our data and those
taken 2 hours earlier by Covino et~al.\ (1999), and the detection in the
second night is not good enough to check its variation over a one-day
period.  The data are consistent with both random fields and ordered field
as sources of the polarized flux. For a random field,
we model the source as consisting of a number of independent
patches, identical in everything but orientation of the field. We then
find that 240--2600 patches are needed to bring the net polarization
down from its intrinsic value of 60\% in each patch to our measured
1.6\%. We also show that future studies of polarization variations can
provide further information about the structure of the magnetic field,
especially about the presence of an ordered component.

\acknowledgements

We are grateful to E. Carretti, 
L. Kaper and V. Radhakrishnan for helpful discussions.


\begin{table}
\begin{minipage}{8.8cm}
   \caption[]{The log of the observations. The observations were 
              performed with the 8.2-m Antu telescope, in standard
              resolution (0.2\arcsec /pixel). 'angle' is the retarder
	      angle, for each standard observation all angles were
	      done (0,22.5,45,67.5).

              \label{tab:log}
	      }
\small
\begin{tabular}{@{}rcccc@{}} \hline
UT date & object &  angle & exposure  & seeing \\
(1999 May) &     &  (deg)         & (s)       & ($^{\prime\prime}$) \\ \hline
11.223   & OT & 0 			& 600  & 1.3 \\
11.231   & OT & 22.5 			& 600  & 1.3 \\
11.239   & OT & 45 			& 600  & 1.4 \\
11.406   & BD--13$^{\circ}$5073 &   	& 0.25 & 0.9 \\
11.425   & BD--12$^{\circ}$5133 &   	& 0.25 & 1.0 \\
12.168   & OT & 0 			& 600  & 2.6 \\
12.175   & OT & 22.5 			& 600  & 2.6 \\
12.183   & OT & 45 			& 600  & 2.6 \\
12.239   & BD--12$^{\circ}$5133 &   	& 0.25 & 2.5 \\
12.249   & BD--13$^{\circ}$5073 &  	& 0.25 & 2.5 \\ \hline
\end{tabular}
\end{minipage}
\end{table}


\end{document}